\documentstyle[12pt]{article}

\catcode`@=11
\def\un#1{\relax\ifmmode\@@underline#1\else
        $\@@underline{\hbox{#1}}$\relax\fi}
\catcode`@=12

\def\a{\alpha}
\def\b{\beta}
\def\c{\chi}
\def\d{\delta}
\def\e{\epsilon}
\def\f{\phi}
\def\g{\gamma}
\def\h{\eta}

\def\j{\psi}

\def\m{\mu}

\def\p{\pi}

\def\s{\sigma}

\def\x{\xi}

\def\D{\Delta}
\def\F{\Phi}
\def\G{\Gamma}

\def\L{\Lambda}

\def\P{\Pi}

\def\U{\Upsilon}
\def\X{\Xi}

\def\ve{\varepsilon}
\def\vf{\varphi}

\def\bo{{\raise-.5ex\hbox{\large$\Box$}}}               
\def\pa{\partial}                                       
\def\pr{\prod}                                          
\def\TH{{\raise.2ex\hbox{$\displaystyle \bigodot$}\mskip-4.7mu \llap H \;}}
\def\face{{\raise.2ex\hbox{$\displaystyle \bigodot$}\mskip-2.2mu \llap {$\ddot
        \smile$}}}                                      

\def\sp#1{{}^{#1}}                              
\def\slash#1{\rlap{\hbox{$\mskip 1 mu /$}}#1}      
\def\Bar#1{\overline{#1}}                       
\def\VEV#1{\left\langle #1\right\rangle}        
\def\leftrightarrowfill{$\mathsurround=0pt \mathord\leftarrow \mkern-6mu
        \cleaders\hbox{$\mkern-2mu \mathord- \mkern-2mu$}\hfill
        \mkern-6mu \mathord\rightarrow$}
\def\dvec#1{\vbox{\ialign{##\crcr
        \leftrightarrowfill\crcr\noalign{\kern-1pt\nointerlineskip}
        $\hfil\displaystyle{#1}\hfil$\crcr}}}           

\def\frac#1#2{{\textstyle{#1\over\vphantom2\smash{\raise.20ex
        \hbox{$\scriptstyle{#2}$}}}}}                   
\def\ha{\frac12}                                        
\def\sfrac#1#2{{\vphantom1\smash{\lower.5ex\hbox{\small$#1$}}\over
        \vphantom1\smash{\raise.4ex\hbox{\small$#2$}}}} 
\def\bfrac#1#2{{\vphantom1\smash{\lower.5ex\hbox{$#1$}}\over
        \vphantom1\smash{\raise.3ex\hbox{$#2$}}}}       
\def\afrac#1#2{{\vphantom1\smash{\lower.5ex\hbox{$#1$}}\over#2}}    

\def\[{\lfloor{\hskip 0.35pt}\!\!\!\lceil}
\def\]{\rfloor{\hskip 0.35pt}\!\!\!\rceil}

\def\fracm#1#2{\hbox{\large{${\frac{{#1}}{{#2}}}$}}}
\def\half{{\fracm12}}
\def\ha{\half}

\def\un{\underline}
\def\fracmm#1#2{{{#1}\over{#2}}}

\def\low#1{{\raise -3pt\hbox{${\hskip 0.75pt}\!_{#1}$}}}

\newskip\humongous \humongous=0pt plus 1000pt minus 1000pt
\def\caja{\mathsurround=0pt}
\def\eqalign#1{\,\vcenter{\openup2\jot \caja
        \ialign{\strut \hfil$\displaystyle{##}$&$
        \displaystyle{{}##}$\hfil\crcr#1\crcr}}\,}
\newif\ifdtup

\def\ref#1{$\sp{#1)}$}

\def\pl#1#2#3{Phys.~Lett.~{\bf {#1}B} (19{#2}) #3}
\def\np#1#2#3{Nucl.~Phys.~{\bf B{#1}} (19{#2}) #3}
\def\prl#1#2#3{Phys.~Rev.~Lett.~{\bf #1} (19{#2}) #3}
\def\pr#1#2#3{Phys.~Rev.~{\bf D{#1}} (19{#2}) #3}
\def\cqg#1#2#3{Class.~and Quantum Grav.~{\bf {#1}} (19{#2}) #3}

\def\mpl#1#2#3{Mod.~Phys.~Lett.~{\bf A{#1}} (19{#2}) #3}

\parskip=\medskipamount                 
\thicklines                         

\def\pc{picture-changing}
\def\pco{picture-changing operator}

\begin{document}


\begin{titlepage}

\noindent
ITP--UH--22/94 \\
hep-th/9412242 \hfill December 1994\\

\vskip 0.6cm

\begin{center}

{\Large\bf From $N{=}2$ Fermionic Strings}\\
\vskip0.5truecm
{\Large\bf to Superstrings?$^{*\,+}$}\\

\vskip 1.0cm

{\large Olaf Lechtenfeld}

\vskip 0.6cm

{\it Institut f\"ur Theoretische Physik, Universit\"at Hannover}\\
{\it Appelstra\ss{}e 2, 30167 Hannover, Germany}\\
{E-mail: lechtenf@itp.uni-hannover.de}\\

\vskip 1cm
\textwidth 6truein
{\bf Abstract}
\end{center}

\begin{quote}
\hspace{\parindent}
{}\ \ \
I review the covariant quantization of the critical $N{=}2$ fermionic string
with and without a global ${\bf Z}_2$ twist.
The BRST analysis yields massless bosonic and fermionic vertex operators
in various ghost and picture number sectors, as well as picture-changers
and their inverses, depending on the field basis chosen for bosonization.
Two distinct GSO projections exist, one (untwisted) retaining merely the
known bosonic scalar and its spectral-flow partner, the other (twisted)
yielding two fermions and one boson, on the massless level. The absence of
interactions in the latter case rules out standard spacetime supersymmetry.
In the untwisted theory, the $U(1,1)$-invariant three-point and vanishing
four-point functions are confirmed at tree level. I comment on the $N{=}2$
string field theory, the integration over moduli and the realization of
spectral flow.

\end{quote}

\vfill

\textwidth 6.5truein
\hrule width 5.cm

{\small
\noindent ${}^*$
Talk at the 28th International Symposium on the
Theory of Elementary Particles,\\
\hspace*{0.4cm}Wendisch-Rietz, Germany, 30 August -- 03 September, 1994 \\
${}^+$
Supported in part by the `Deutsche Forschungsgemeinschaft'
}

\eject
\end{titlepage}

\newpage
\hfuzz=10pt

In this talk I am reporting on some recent progress in unraveling
the BRST structure of the critical $N{=}2$ fermionic string.
This investigation has been a joint effort including Jan Bischoff,
a student of mine, Sergei Ketov, postdoc at Hannover, as well as
Andrew Parkes, now at Edinburgh. For more details, I refer to our
preprints~\cite{klp,bkl}.

There are a number of reasons why the $N{=}2$ extended fermionic
string~\cite{aba,aba2} is worth studying in spite of the fact that
its critical real dimension equals $4{+}0$ or $2{+}2$,
excluding $D=1{+}3$~\cite{ft,ali}.
First, there is the curious fact that this object is actually just a
point particle pretending to be a string, since it has only a finite
number of physical excitations~\cite{bien}.
Second, it is natural to wonder if the known connection between
world-sheet and spacetime supersymmetry extends to this case, as
has been advocated by Siegel~\cite{siegel}.
Third, $N{=}2$ strings are known to be closely related with self-dual
four-dimensional field theories and integrable models~\cite{ov}.
Fourth, the $N{=}0$ (bosonic) and $N{=}1$ (fermionic) strings can also
be considered in the framework of the general $N{=}2$ string theory as
particular vacua~\cite{bv,ff,op,bop}.
Fifth, the moduli spaces of $N{=}2$ super Riemann surfaces carry novel
structure due to the presence of an additional $U(1)$ gauge field.
Finally, the mere existence of this string theory warrants its study.
For a review of the subject, consult refs.~\cite{ma,ket}.

We begin by reviewing the BRST quantization of the $N{=}2$ string and
add new results concerning possible twisting, chiral bosonization, and
picture-changing and its inverse, as we go along.
Our starting point is the $N{=}2$ world-sheet supergravity action~\cite{bsa}.
The extended supergravity multiplet involves the real zweibein $e^a_\a$,
a {\it complex\/} gravitino $\c_\a$ and a real abelian gauge field
(graviphoton) $A_\a$, with $\a=0,1$.
The conformal matter consists of two complex string coordinates $Z^\m$ and
two complex spin $1/2$ (NSR) fermions $\j^\m$, with $\m=0,1$.
In this talk, I choose the signature of the real world-sheet metric and
the complex spacetime metric both to be $(-+)$.

When decomposing the complex matter degrees of freedom into real ones,
two different field bases suggest themselves:
A `real basis' $Z^{i\m}$, $i=2,3$, should read
$$
(Z^{2\m},Z^{3\m}):=({\rm Re} Z^\m,{\rm Im} Z^\m) \quad,
\eqno(1a) $$
whereas a `holomorphic basis' would be
$$
(Z^{+\m},Z^{-\m}):=(Z^\m,Z^{*\m}) \quad.
\eqno(1b) $$
The unusual range of the index~$i$ helps to avoid confusing the same
numerical values of $\m$ and~$i$.

The coordinates $Z$ take values in the target space ${\bf C}^{1,1}$.
Besides the local $N{=}2$ superconformal world-sheet symmetry,
the action also has a global spacetime symmetry given by rigid translations
plus `Lorentz transformations' comprising $U(1,1)\times{\bf Z}_2$~\cite{ov}.
The ${\bf Z}_2$ factor amounts to complex conjugation.
The string can be twisted by identifying fields upon complex conjugation,
e.g. $Z^\m\sim Z^{*\m}$.
In the real basis, this effectively puts $Z^{3\m}=0$ and changes the target
to the half-spacetime ${\bf C}^{1,1}/{\bf Z}_2$.

Transporting the matter fields along a non-contractible cycle on the
world-sheet, the NSR fermions $\j^{i\m}$ show a ${\bf Z}_2$
monodromy, since they live in a spin bundle.~\footnote{
In the untwisted case, the monodromy group is actually $U(1)$ due to the
spectral flow in the $N{=}2$ superconformal algebra.
We shall comment on this later.}
For the bosonic coordinates $Z^{i\m}$ the monodromy group is either trivial
(untwisted theory) or again ${\bf Z}_2$ (twisted case), depending on whether
or not we allow $Z^{3\m}$ to be antiperiodic.
`Lorentz invariance' demands the monodromies to be independent of the
$\m$~index. Moreover, single-valuedness of the Brink-Schwarz lagrangian
requires that the product of the $(i{=}2)$ and $(i{=}3)$ monodromies be the
same for $Z$ and for $\j$. Thus, one ends up with four sectors:

\begin{center}
\noindent\begin{tabular}{c|cc|cc}
sector & $Z^{2\m}$ & $Z^{3\m}$ & $\j^{2\m}$ & $\j^{3\m}$ \\
\hline
(NS,NS) & P & P & P & P \\
(R,R)   & P & P & A & A \\
\hline
(NS,R)  & P & A & P & A \\
(R,NS)  & P & A & A & P \\
\end{tabular}
\end{center}

\noindent
Here, P and A refer to periodic and antiperiodic boundary conditions,
respectively, in natural coordinates on the cylinder.
The untwisted string consists of the first two sectors only,
which then are connected by spectral flow~\cite{ov}.
The twisted theory contains all four sectors and was first considered
by Mathur and Mukhi~\cite{mm}.
As a consequence of the twist, however, the graviphoton $A_\a$ becomes
antiperiodic and, hence, must be set to zero.
This means that the $N{=}2$ world-sheet supersymmetry gets broken to $N{=}1$
by the ${\bf Z}_2$ twist, and the spectral flow disappears in this case.
More general monodromies from $U(1,1)$ are compatible with the action,
but have been shown not to lead to massless physical states (with one
curious exception)~\cite{klp}.

Via BRST quantization in the $N{=}2$ superconformal gauge the fields (and
ghosts) of the $N{=}2$ string on the euclidean world-sheet become free, so
that they can be decomposed into their holomorphic and anti-holomorphic parts,
as is usual in 2d conformal field theory. From now on, our notation will
refer to the left-moving (holomorphic) part of the conformal fields.
The ghost systems appropriate for the $N{=}2$ string are:
\begin{itemize}
\item the reparametrization ghosts ($b,c$), an anticommuting pair of
free world-sheet fermions with conformal dimensions~($2,-1$).
\item the two-dimensional $N{=}2$ supersymmetry ghosts ($\b^i,\g^i$)
or ($\b^\mp,\g^\pm$), two commuting pairs of free world-sheet fermions
with conformal dimensions~($\frac32,-\frac12$).
\item the $U(1)$ ghosts ($\tilde{b},\tilde{c}$), an anticommuting pair of
free world-sheet fermions with conformal dimensions~($1,0$).
\end{itemize}

The $N{=}2$ string BRST charge
$$Q_{\rm BRST}\ =\ \oint_0 \fracmm{dz}{2\p i}\,j_{\rm BRST}(z)
\eqno(2) $$
is expressed through the dimension-one BRST current which in its most
convenient form reads
$$\eqalign{
j_{\rm BRST}\ =&\ c\hat{T} + bc\pa c +\g^2G +\g^3{\Bar G}+\tilde{c}\hat{J}\cr
&-(\g^2\g^2 + \g^3\g^3)b +2i(\g^2\pa\g^3 - \g^3\pa\g^2)\tilde{b}
+\fracm34 \pa[c(\b^2\g^2 +\b^3\g^3)]\cr}\eqno(3a) $$
in the real basis, or
$$\eqalign{
j_{\rm BRST}\ =&\ c\hat{T} + bc\pa c + \fracm12\g^-G^+ + \fracm12\g^+G^-
+\tilde{c}\hat{J}\cr
&- \g^+\g^-b + (\g^-\pa\g^+ -\g^+\pa\g^-)\tilde{b}
+\fracm38 \pa[c(\b^+\g^- +\b^-\g^+)]\cr}\eqno(3b) $$
in its holomorphic form.
Here, we use the notation
$$\hat{T}\ =\ T_{\rm tot}-T_{b,c}\qquad{\rm and}\qquad
\hat{J}\ =\ J_{\rm tot}-\pa(\tilde{b}c)~,\eqno(4) $$
where $T_{b,c}=-2b\pa c -(\pa b)c$, and we introduced the full (BRST-invariant)
stress tensor $T_{\rm tot}$ and the $U(1)$ current $J_{\rm tot}$ as
$$\eqalign{
T_{\rm tot}\ =\ \{Q_{\rm BRST},b\}\ &=\
T+T_{b,c} -\tilde{b}\pa\tilde{c}-\fracm32
(\b^2\pa\g^2+\b^3\pa\g^3)-\fracm12(\g^2\pa\b^2+\g^3\pa\b^3)\cr
&=\ T+T_{b,c} -\tilde{b}\pa\tilde{c}-\fracm34
(\b^+\pa\g^-+\b^-\pa\g^+)-\fracm14(\g^+\pa\b^-+\g^-\pa\b^+)~,\cr
J_{\rm tot}\ =\ \{Q_{\rm BRST},\tilde{b}\}\ &=\
J +\pa(\tilde{b}c)-\fracm{i}2(\b^2\g^3-\b^3\g^2)\cr
&=\ J +\pa(\tilde{b}c)+\fracm14(\b^+\g^--\b^-\g^+)~.\cr}
\eqno(5) $$
Above, $T$, $G$ and $J$ are the $N{=}2$ string (matter) currents
without ghosts, {\it viz.}
$$\eqalign{
T\ &=\ -\ha\left(\pa Z^i\cdot\pa Z^i - \j^i\cdot\pa\j^i\right)\
=\ -\ha\pa Z^+\cdot\pa Z^-
+\fracm14\j^+\cdot\pa\j^-+\fracm14\j^-\cdot\pa\j^+~,\cr
G\ &=\ \d^{ij}\pa Z^i\cdot\j^j \qquad,\qquad\qquad\qquad\qquad
G^+\ =\ \pa Z^-\cdot\j^+ ~,\cr
{\Bar G}\ &=\ \ve^{ij}\pa Z^i\cdot\j^j \qquad,\qquad\qquad\qquad\qquad
G^-\ =\ \pa Z^+\cdot\j^- ~,\cr
J\ &=\ \fracm{i}{4}\ve^{ij}\j^i\cdot\j^j\
=\ \fracm{i}{2}\j^2\cdot\j^3 \
=\ -\fracm14 \j^+\cdot\j^-~.\cr} \eqno(6) $$

In order to construct fermionic vertex operators we make use of chiral
bosonization~\cite{fms,k-w}.
This requires us to switch to a Cartan-Weyl or light-cone basis with
respect to the index $\m$, e.g.
$$
\j^{\cdot\pm}\ =\ \j^{\cdot 0}\pm\j^{\cdot 1} \quad.
\eqno(7) $$
For the complex fields $\j$, $\b$, and $\g$, bosonization depends on
the basis. As for the $N{=}1$ string, the Fock space of the first-order
fields gets imbedded into an extended Fock space of second-order bosonic
fields,
$$
\eqalign{
{\rm real~basis:}\qquad\qquad\qquad\quad
\j^{i+}, \j^{i-}; \b^i, \g^i \quad &\Longrightarrow\quad
\f^i; \vf^i, \h^i, \pa\x^i \cr
{\rm holomorphic~basis:}\qquad
\j^{\pm+}, \j^{\pm-}; \b^\pm, \g^\pm\quad &\Longrightarrow\quad
\f^\pm; \vf^\pm, \h^\pm, \pa\x^\pm \quad, \cr}
\eqno(8) $$
where we chose not to bosonize the auxiliary fermionic $(\h,\x)$ system.
The creation of Ramond states out of the Neveu-Schwarz vacuum requires the
use of matter and ghost spin fields, which are constructed as
$e^{\pm\frac12\f}$ and $e^{\pm\frac12\vf}$, respectively. Furthermore,
matter and ghost twist fields are needed to generate twisted states.
It should be stressed that the relation between the two bosonization schemes
is non-local and non-polynomial, although the unbosonized field bases are
linearly related. This is exemplified by
$$\eqalign{
\hat J\ &=\
\fracm{i}{2} \j^2{\cdot}\j^3 + \fracm{i}{2} (\g^2\b^3-\g^3\b^2) \ =\
-\fracm{i}{2} e^{+\f^2-\f^3} + \fracm{i}{2} e^{+\vf^2-\vf^3} \h^2\pa\x^3
- (2\leftrightarrow 3) \cr
&=\ -\fracm14 \j^+{\cdot}\j^- - \fracm14 (\g^+\b^- -\g^-\b^+) \ =\
\fracm12 (\pa\f^+ -\pa\f^-) + \fracm12 (\pa\vf^+ -\pa\vf^-)~.\cr}
\eqno(9) $$

The BRST cohomology problem is simplified by identifying grading operators.
For the $N{=}2$ string, these are
\begin{itemize}
\item the total ghost charge
$U=-\oint[bc+\tilde b\tilde c+\fracm12\b^+\g^-+\fracm12\b^-\g^+]$
\item the picture charges
$\ \Pi^i=-\oint[\b^i\g^i+\h^i\x^i]$\quad\qquad\qquad $i=2$ or $3$ \\
\phantom{the picture char} or
$\ \Pi^\pm=-\fracm12\oint[\b^\pm\g^\mp+\h^\pm\x^\mp]$\quad\qquad $+$ or $-$
\item the full bosonic constraints $\ T_{\rm tot}$ and $J_{\rm tot}$
\end{itemize}
of which only $U$ does not commute with $Q_{\rm BRST}$.
Accordingly, it suffices to separately investigate simultaneous eigenspaces
of the commuting set $\{U,\Pi,L_0^{\rm tot},J_0^{\rm tot}\}$, labelled
by $\{u,\pi,h,e\}$. Note that the picture charges are basis-dependent. From
$L_0^{\rm tot}=\{Q_{\rm BRST},b_0\}$ and
$J_0^{\rm tot}=\{Q_{\rm BRST},\tilde b_0\}$
it readily follows that non-trivial cohomology only exists
for $h=e=0$.~\footnote{
Note, however, that $J_0^{\rm tot}$ and $e$ are not defined for twisted
states.}
One can show that the total ghost number $u$ takes on integral values, while
the two picture numbers $\p$ may each be integral (NS) or half-integral (R),
corresponding to the four choices of the NSR monodromies.
Obviously, $\p^++\p^-\in{\bf Z}$ for the untwisted string.
No further restrictions on the values of $u$ or $\p$ arise at this point,
so that an infinity of (massless) physical vertex operators is anticipated.
Like in the $N{=}1$ string, however, BRST cohomology classes differing by
integral values of $u$ or $\p$ should correspond to the same physical state, if
their spacetime properties agree. Hence, we should identify physical states
with equivalence classes of BRST cohomology classes under integral total ghost
and picture number changes.

In ref.~\cite{bkl} we have investigated the massless BRST cohomology for
various
total ghost numbers and generic spacetime momenta in the pictures between
$({-}2,{-}2)$ and $(0,0)$, in the real as well as the holomorphic basis.
Given the picture, we found four non-trivial untwisted classes in
three ghost sectors, and four non-trivial twisted classes in two ghost sectors.
When restricting the lightlike spacetime momentum according to the
twisted target spacetime, $k^{3\m}=0$, each untwisted class splits into two,
a spacetime left-mover ($k^{2+}=0$, labelled with `$+$' superscript)
and a spacetime right-mover ($k^{2-}=0$, labelled with `$-$' superscript).
The non-trivial ghost sectors, together with their multiplicity, are given by
$$
v\ \equiv\ u-\p^2-\p^3\ =\
1^\pm, \fracm32^\pm, 2^\pm, 2'^\pm, \fracm52^\pm, 3^\pm
\eqno(10a) $$
for the twisted string and by
$$
v\ \equiv\ u-\p^+-\p^-\ =\ 1, 2, 2', 3
\eqno(10b) $$
in the untwisted string,
where the superscripts distinguish the distinct classes.
We conjecture that this pattern persists for all pictures,
and that no further physical states appear at massive levels.
Then, the twisted string (in real basis) has the following physical spectrum
for restricted lightlike spacetime momenta $k^2\cdot k^2=0$, $k^{3\m}=0$:

\begin{center}
\noindent\begin{tabular}{c|c|c}
sector & ground states & statistics \\
\hline
(NS,NS) & $\F^+\quad\F^-$ & even \\
(R,R)   & $\U^+\quad\U^-$ & even \\
(NS,R)  & $\X^+\quad\X^-$ & odd  \\
(R,NS)  & $\L^+\quad\L^-$ & odd  \\
\end{tabular}
\end{center}

\noindent
In contrast, the massless and neutral ground states of the untwisted string are
a single $\F$ in the (NS,NS) sector and a single $\U$ in the (R,R) sector,
for generic momenta $k^{i\m}$ satisfying $k^2\cdot k^2+k^3\cdot k^3=0$.
The form of the vertex operators depends on the field basis.
We shall argue that $\F$ and $\U$ are related by spectral flow.

The individual ghost number selection rules for tree-level amplitudes demand
that non-vanishing correlation functions have overall total ghost and
picture charges of zero and minus two, respectively, and, hence,
$v_{\rm total}=4$. Of course, spacetime momenta have to add to zero.
The non-vanishing two-point functions yield a pairing of vertex operators with
$$
\bigl(v;\p^+,\p^-;k^{\pm\m}\bigr)\quad \longleftrightarrow\quad
\bigl(4{-}v;-2{-}\p^+,-2{-}\p^-;-k^{\pm\m}\bigr)
\eqno(11) $$
in the holomorphic basis and similarly for the real basis.
Setting those two-point functions equal to one normalizes the vertex operators.
Clearly, the canonical choice for $\F$ is the (${-}1,{-}1$) picture,
with
$$
V_\F\ =\ V^{(1)}_{(-1,-1)}\ =\ c\,e^{-\vf^--\vf^+}\,e^{ik\cdot Z}
\eqno(12) $$
for $k\cdot k=0$ and $v=1$ (indicated as superscript) in the holomorphic basis.
For $\U$ the situation is asymmetric. Here, the canonical vertex operator is
$$
\eqalign{
V_\U\ &=\ V^{(1)}_{(-\frac32,-\frac32)}\ =\
c\,e^{-\frac32\vf^--\frac32\vf^+}\,h_\a (S^+S^-)^\a\,e^{ik\cdot Z} \cr
{\rm or}\ &=\ V^{(1)}_{(-\frac12,-\frac12)}\ =\
c\,e^{-\frac12\vf^--\frac12\vf^+}\,h_\a (\slash{k}^+\slash{k}^-S^+S^-)^\a\,
e^{ik\cdot Z} ~,\cr}
\eqno(13) $$
with the spin field product
$$
(S^+S^-)^\pm\ =\ e^{\pm\frac12\f^+\pm\frac12\f^-}
\eqno(14) $$
being an $SO(2,2)$ Weyl spinor (with $SU(1,1)$ spinor index $\a=\pm$).
The polarization spinor $h_\a$ satisfying
$h_\a k^{\a\dot\a}=0$~\footnote{
The $SO(2,2)$ vector $k^{i\m}$ is written as an $SU(1,1)$ bispinor.}
represents a single degree of freedom since it must be proportional either
to $k^{\a+}$ or to $k^{\a-}$.
The two resulting forms of the vertex, $V^+_\U$ and $V^-_\U$,
are BRST equivalent for generic $2{+}2$ dimensional momenta.
In the twisted case the $\U$ cohomology splits into $\U^+$ and $\U^-$.
It follows that
$$
\VEV{V^{(1)}_{(-\frac32,-\frac32)}(k,h)\quad
V^{(3)}_{(-\frac12,-\frac12)}(-k,\bar h)}\ =\
h_a\,(\slash{k}^+\slash{k}^-)^{\a\b}\,\bar h_\b\
\mathrel{\mathop=^!} 1
\eqno(15) $$
which becomes singular when some light-cone projection $k^{\cdot\pm}$ vanishes.
An extensive list of explicit vertex operators can be found in the appendix
of ref.~\cite{bkl}.

Fortunately, we do not need to repeat the cohomology analysis for each picture,
since an explicit equivalence relation is known. More precisely,
the so-called \pc\ operations~$X^i$~\cite{nst,fms}
shift $\pi^j\to\pi^j+\d^{ij}$ and $u\to u+1$
while commuting with $Q_{\rm BRST}$, $b_0$ and~$L_0^{\rm tot}$,
and leaving $v$ unchanged.~\footnote{
The modified ghost number $v$ was conveniently introduced for this reason.}
Picture-changing in the real basis proceeds as
$$\eqalign{
V_{(\pi^2+1,\pi^3)}\ =&\ [Q_{\rm BRST},\x^2 V_{(\pi^2,\pi^3)}\}\ =:\
X^2\cdot V_{(\pi^2,\pi^3)} \cr
V_{(\pi^2,\pi^3+1)}\ =&\ [Q_{\rm BRST},\x^3 V_{(\pi^2,\pi^3)}\}\ =:\
X^3\cdot V_{(\pi^2,\pi^3)} ~,\cr}\eqno(16) $$
where, in the absence of normal ordering between $\x^i$ and $V$, one defines
the real \pco s
$$ X^i(z)\ :=\ \{Q_{\rm BRST},\x^i(z)\}\qquad\qquad i=2,3~.\eqno(17a) $$
Analogously, in the holomorphic basis,
$$ X^\pm(z)\ :=\ \{Q_{\rm BRST},\x^\pm(z)\} \eqno(17b) $$
shifts $\p^+$ or $\p^-$ by one unit, while leaving
$v=u-\p^+-\p^-$ untouched.

By construction, the $X$ are BRST invariant but {\it not} BRST trivial
due to the lack
of the zero modes of $\x$ in the bosonization formulae~\cite{fms}.
The \pco s just introduced take the following explicit form:
$$\eqalign{
X^2\ =~&~c\pa\x^2 +e^{\vf^2}\left[ G + \fracm{i}{2}\tilde{c}\b^3
 + 2i(\pa\tilde{b})\g^3 + 4i\tilde{b}\pa\g^3\right]
+2e^{2\vf^2}b\pa\h^2 +\pa(e^{2\vf^2}b)\h^2 ~,\cr
X^3\ =~&~c\pa\x^3 +e^{\vf^3}\left[ {\Bar G} - \fracm{i}{2}\tilde{c}\b^2
 - 2i(\pa\tilde{b})\g^2 - 4i\tilde{b}\pa\g^2\right]
 + 2e^{2\vf^3}b\pa\h^3 +\pa(e^{2\vf^3}b)\h^3 ~,\cr}\eqno(18a) $$
$$\eqalign{
X^+\ =~&~c\pa\x^+ +e^{\vf^-}\left[ G^+
+2\pa\tilde{b}\g^+ +4\tilde{b}\pa\g^+ -2b\g^+ \right]~,\cr
X^-\ =~&~c\pa\x^- +e^{\vf^+}\left[ G^-
-2\pa\tilde{b}\g^- -4\tilde{b}\pa\g^- -2b\g^- \right] ~.\cr}\eqno(18b) $$
It is clear that $X^\pm$ are not just linear combinations of $X^i$.
The two types of \pco s differ in two respects.
The holomorphic version does not contain $e^{2\vf}$ terms, and it also lacks
any $\tilde{c}$ dependence. The latter means that $X^\pm$ commute with
$J_0^{\rm tot}$, whereas $X^i$ do so only modulo BRST-exact terms.

Since \pc\ $X$ establishes an equivalence of cohomology classes,
its inverse $Y$ can only be well-defined modulo BRST-trivial terms and may,
like $X$ itself, possess a BRST-trivial nonzero kernel. We require in the
real form that
$$ [Q_{\rm BRST},Y^i]\ =\ 0 \qquad\qquad{\rm and}\qquad\qquad
Y^2(z)\;X^2(w)\ \sim\ 1\ \sim\ Y^3(z)\;X^3(w) \eqno(19) $$
but do not constrain the mixed products.
The quantum numbers of $Y^i$ are determined as
$(h,u,\p^j)=(0,-1,-\d^{ij})$. A simple ghost number analysis shows that this
leaves only a single candidate for each $\vf^i$ ghost charge value below $-1$.
In the case of $Y^2$, for instance, we are forced to write a
linear combination of
$$ Y^2_k\ =\ c\,\left(\g^3\right)^{k-2}\,\pa^{k-1}\x^2\ldots\pa^2\x^2\pa\x^2\,
e^{-k\vf^2} \qquad\quad k\ge2 \eqno(20) $$
which satisfy
$$ Y^2_k(z)\,X^2(w)\ \sim\ \d_{k2} + O(z{-}w) ~.\eqno(21) $$
The promising first term,
$$ Y^2_2\ =\ c\,\pa\x^2\,e^{-2\vf^2}~,\eqno(22) $$
is identical with the inverse picture-changing operator of the $N{=}1$ string,
but fails to be BRST invariant for the $N{=}2$ string.
Astoundingly, this failure can be corrected by adding an infinite series of
$Y^2_k$, with $k=4,6,8,\ldots$. In other words, the coefficients in
$$ Y^2\ =\ \sum_{{k=2\atop{k~\rm even}}}^\infty
\left(\prod_{\ell=1}^{k-1}\ell!\right)^{-1} Y^2_k \ \ =\ \
Y^2_2+{1\over12}Y^2_4+{1\over34560}Y^2_6+\ldots\eqno(23) $$
lead to a chain of cancellations among BRST commutators of successive terms.
With some effort, the formal series can be summed to the non-local expression
$$ Y^2(w)\ =\ \sin \oint_w [\g^2\b^3-\g^3\b^2] \cdot Y^2_1(w) \ =\
-i \sinh (2~{\rm ad}J_0^{\rm tot})\cdot Y^2_1(w) \eqno(24) $$
where we introduced
$$ Y^2_1\ =\ -c\,\x^3\,e^{-\vf^3-\vf^2}\ =\ c(\g^3)^{-1}\d(\g^2) \eqno(25) $$
and understand the action on $Y^2_1$ as a power series of iterated commutators.
Of course, a mirror image expression emerges for~$Y^3$.
Note that the $Y^i$ are pure ghost operators and do not contain
any matter fields.
A similar analysis in the holomorphic basis fails to produce any candidate
for~$Y^\pm$. Still, we suspect that some, necessarily non-local, inverse \pco s
exist in this case as well.

So far, our treatment of local vertex operators has not been systematic.
Now we are going to employ a unified formalism which
simultaneously deals with all chiral vertex operators defined in arbitrary
pictures. For the purpose of mutual locality of vertex operators, we may
temporarily forget about their momentum-dependence~\footnote{
For the tree-level correlation functions, their momentum dependence is
essentially absorbed into the usual Koba-Nielsen factor.
In the twisted sector, the constrained kinematics (only $k^{2-}$ non-zero)
allows merely $Z^{2+}$ which does not get twisted.}
and concentrate on their ghost, spin, and twist field structure.
In operator products,
the $(b,c)$ and matter twist fields never lead to branch cuts,
since $t^3$ always occurs in combination with a ghost twist field, i.e.
$t^3_+=t^3 e^{\tilde\s/2}$, and those have meromorphic OPE with one another.
Only $\j^{i\pm}$ and $S^{i\pm}$ as well as $(\b^i,\g^i)$ and their spin
fields may ruin locality.  After real bosonization,
any vertex operator is a linear combination of terms proportional to~\footnote{
Eventual derivatives of $\f^i$ or $\vf^i$ are irrelevant again.}
$$
\exp\left[p_2\f^2 + q_2\vf^2 + p_3\f^3 + q_3\vf^3\right]~,
\eqno(26) $$
where $p_i$ and $q_i$ take integral or half-integral values.

More specifically, each pair $[p_i,q_i]$ belongs to a lattice
$$
\G_w^{(i)}\ =\ \P^{1,1}\ \equiv\
{\bf Z}^{1,1}\cup\bigl[(\ha,\ha)+{\bf Z}^{1,1}\bigr]\
=\ \underbrace{(o) \cup (v)}_{{\rm NS}} \cup
\underbrace{(s) \cup (c)}_{{\rm R}}
\eqno(27) $$
which may be regarded as a half-integral lorentzian weight lattice.
Here, the scalar product has been chosen as $[p,q]\cdot[p',q']=pp'-qq'$.
Like for $so(2n)$,
the weight lattice decomposes into the root lattice $(o)$ and three copies
of it, each shifted by a different elementary weight vector and conventionally
denoted by $(v)$, $(s)$ and $(c)$.
The lorentzian length-squared $p^2-q^2$ is even integer except for weights
in~$(v)$ where it is odd.
On the other hand, the contribution to the conformal dimension
$h=\ha\sum_i(p_i^2-q_i^2-2q_i)$ of the
operator~(26) is integral for~$(o)$ and
half-integral for $(v)$, $(s)$ and $(c)$.

For our purposes, we have to consider the combined weights
$[p_2,q_2;p_3,q_3]$ which form the still half-integral $2{+}2$ dimensional
weight lattice
$$
\G_w\ =\ \G_w^{(2)}\oplus\G_w^{(3)}\ =\ \bigl\{(r_2;r_3)\bigr\}~.
\eqno(28) $$
{}From the 16 conjugacy classes $(r_2;r_3)$ those 6 containing a
single $(o)$ contribute half-integrally to $h$ and, hence, do not contain
physical states.
The remaining 10 classes $\G_w'$ do not form a lattice.
They split into 6 {\it even\/} (= untwisted) and
4 {\it odd\/} (= twisted) classes:

\begin{center}
\noindent\begin{tabular}{c|cccccc|cccc}
class & $(o,o)$ & $(v,v)$ & $(s,s)$ & $(c,c)$ & $(s,c)$ & $(c,s)$ &
$(v,s)$ & $(v,c)$ & $(s,v)$ & $(c,v)$ \cr
\hline
state & -- & $\F^\pm$ & $\U^-$ & $\U^+$ & -- & -- &
$\X^-$ & $\X^+$ & $\L^-$ & $\L^+$ \cr
\end{tabular}
\end{center}

\noindent
Like in the $N{=}1$ string, we should like to identify commuting
(spacetime bosonic) vertex operators with {\it even untwisted\/} weights
and anticommuting (spacetime fermionic) vertex operators with
{\it odd twisted\/} weights.
The lattice consideration takes us out of the `canonical pictures'
because the fusion algebra only closes in the infinite set of all pictures.
Since the picture numbers $\p^i$ agree
{\it modulo 1\/} with~$q_i$, integral (half-integral) values of
$\p^2+\p^3$ coincide with even untwisted bosonic (odd twisted fermionic)
vertex operators, and we may use these terms interchangeably.

The relevance of the lattice description derives from the
basic OPE of two bosonized operators as in eq.~(26),
$$
\exp[p_i\f^i +q_i\vf^i](z)\,\exp[p'_i\f^i +q'_i\vf^i](w)
\sim (z-w)^{p_ip'_i - q_iq'_i} \exp[(p_i+p'_i)\f^i +(q_i+q'_i)\vf^i](w)~,
\eqno(29) $$
which relates the mutual locality of two vertex operators to the integrality
of their weight's lorentzian scalar product.
Moreover, fusion simply corresponds to adding weights.
Our task is to enumerate all local vertex operator subalgebras.
Apparently, this amounts to classifying the {\it integral sublattices\/}
$\G_{\rm int}\subset\G_w'$.
Each such $\G_{\rm int}$ is obtained from $\G_w$
by some GSO projection~\cite{gos} and leads to a different string model.
By inspection, one finds six maximal possibilities,
$$\eqalign{
I\qquad\qquad    & (o;o)\ \cup\ (v;v)\ \cup\ (s;s)\ \cup\ (c;c) \cr
II\qquad\qquad   & (o;o)\ \cup\ (v;v)\ \cup\ (s;c)\ \cup\ (c;s) \cr
III\qquad\qquad  & (o;o)\ \cup\ (s;s)\ \cup\ (v;c)\ \cup\ (c;v) \cr
IV\qquad\qquad   & (o;o)\ \cup\ (c;c)\ \cup\ (v;s)\ \cup\ (s;v) \cr
V\qquad\qquad    & (o;o)\ \cup\ (s;c)\ \cup\ (v;s)\ \cup\ (c;v) \cr
VI\qquad\qquad   & (o;o)\ \cup\ (c;s)\ \cup\ (v;c)\ \cup\ (s;v) ~.\cr}
\eqno(30) $$
By construction, these lattices are self-dual, which is expected to be crucial
for modular invariance.
It is non-trivial that these projections are compatible with picture-changing,
since $X^i\in(o,o)\cup(v,v)$ and $Y^i\in(o,o)$.
Helicity flips $(s)\leftrightarrow(c)$
connect $I$ and~$II$ as well as $III$ through~$VI$.
Hence, there are only {\it two\/} types of essentially distinct
GSO projections, say, model~$I$, with four bosonic classes,
and model~$III$, with two bosonic and two fermionic ones.
We may call them `notwist' and `twisted', respectively.
Clearly, the surviving physical states are
$$\eqalign{
{\rm GSO}_{\rm notwist}\quad &\Longrightarrow\quad (\F,\U) \quad,\cr
{\rm GSO}_{\rm twisted}\quad &\Longrightarrow\quad (\U^-,\X^+,\L^+)\quad.\cr}
\eqno(31) $$
{}From the analysis
of the type~III GSO projection in eq.~(30), we should expect a {\it second\/}
bosonic state in~$(o;o)$ in addition to $\U^-\in(s;s)$. Using picture
equivalence, this new state must be represented in the $({-}1,{-}1)$ picture.
However, dimensional analysis easily shows that there can be no massless
$(o;o)$ state in this picture; the only massless state at all is~$\F\in(v;v)$!
We must conclude that for the twisted theory the $(o;o)$ class has no BRST
cohomology, i.e. contains only trivial states.
Unfortunately, there is no match between bosons and fermions.

Let us take a closer look at the twisted model since it holds the
promise of spacetime supersymmetry.
The twisted spacetime momentum constraint $k^{3\m}=0$ implies for
lightlike momenta that the two fermions are left-movers ($k^{2+}=0$)
while the boson is a right-mover ($k^{2-}=0$).
Effectively, the target space dimensionality has been reduced to $D=1{+}1$,
excluding interactions between right- and left-movers.
Indeed, it is not hard to check that the fusion rules are trivial,
i.e. any two fields fuse to a BRST trivial state.
Consequently, the only allowed three-point function vanishes,
$\VEV{\U^-\X^+\L^+}=0$.
The prospective spacetime supersymmetry generators,
$$
Q^2\ =\ \oint b_{-1}\,V_{\L^+}(k{=}0) \qquad\qquad
Q^3\ =\ \oint b_{-1}\,V_{\X^+}(k{=}0) \quad,
\eqno(32) $$
satisfy
$$
\bigl\{ Q^i\,,\,Q^j\bigr\}\ =\ 0 \qquad\quad i,j=2,3 \quad.
\eqno(33) $$
They qualify as exterior derivatives rather than supersymmetry charges.

The untwisted string, in contrast, permits interactions.
Its fusion rules read
$$
[\F]\cdot[\F]\ =\ [\F]\quad,\qquad
[\F]\cdot[\U]\ =\ [\U]\quad,\qquad
[\U]\cdot[\U]\ =\ [\F]\quad,
\eqno(34) $$
and the tree-level three-point functions become
$$
\VEV{V_\F(k_1)\,V_\F(k_2)\,V_\F(k_3)}\ =\
\VEV{V_\F(k_1)\,V_{\U}(k_2)\,V_{\U}(k_3)}\ =\
\e^{ij}\,k_2^i \cdot k_3^j ~,
\eqno(35) $$
the only non-zero bilinear $U(1,1)$ invariant.
All possible four-point functions have been checked to vanish identically,
due to non-trivial kinematic identities in $2{+}2$ dimensions~\cite{ov}.
It has been conjectured that all higher-point functions and loop correlators
vanish as well.
If this is correct, the effective spacetime action for the $\F$ degree of
freedom will be that of self-dual Yang-Mills for the open-string case or
of self-dual gravity for the closed-string case, in $2{+}2$ dimensions.

Like in the $N{=}1$ superstring, \pco s have singular OPE with one another,
$$
X^i(z)\,X^j(w)\ \sim\ (z{-}w)^{-2}\,\{ Q_{\rm BRST},\D^{ij} \}
\qquad i,j=2,3 \quad,
\eqno(36) $$
which spoils the gauge invariance of Witten's superstring field theory
with the NS string field in the $-1$ picture~\cite{wi,we,ls1}.
In the holomorphic basis of the $N{=}2$ string, however, {\it like\/}
\pco s show a {\it regular\/} OPE.
Thus, we can follow the recipe given in ref.~\cite{ls2} and use left-moving
$X^+$s and right-moving $X^-$s exclusively, which avoids singular $X$
collisions anywhere in the moduli space.
We conclude that the `canonical' $N{=}2$ string field theory is well-behaved
and does not suffer from singular contact terms.

Let me close with a few remarks on the structure of loop amplitudes.
There have been some preliminary one-loop computations~\cite{mm,bgi}
but it is fair to say that the issue is still unclear.
A closed $N{=}2$ super Riemann surface is characterized
by its genus~$g\in{\bf N}$
and its instanton or Chern number~$c\in{\bf Z}$
as well as (in complex count)
\begin{itemize}
\item $3g{-}3$ metric moduli,
\item $2g{-}2{+}c$ positively charged fermionic moduli,
\item $2g{-}2{-}c$ negatively charged fermionic moduli,
\item $g$ abelian gauge or $U(1)$ moduli,
\end{itemize}
for $g>1$.
The string measure supports only a non-negative number of fermionic moduli,
which restricts the sum over~$c$ for any given~$g$.

The metric moduli are well studied, and the fermionic moduli will have to
be integrated out formally around split surfaces.
The abelian gauge moduli are a new feature of the $N{=}2$ string.
The $U(1)$ moduli space is the space of flat $U(1)$ connections which
is nothing but the Jacobian torus $J={\bf C}^g/({\bf Z}^g+\Omega{\bf Z}^g)$.
Since any flat $U(1)$ connection $A$ is characterized by the $2g$ phases
$\exp{\oint A}$ around the homology cycles, there is a one-to-one
correspondence between $J$ and the complex torus $Pic={\bf C}^g/{\bf Z}^{2g}$
which describes the space of holomorphic line bundles with a given degree.
A point in $Pic$ determines the twists on the homology, i.e. the constant
monodromy phases which are the transition functions around the cycles.
The spin structures labelling holomorphic spinor bundles sit on the half-points
$(\frac12{\bf Z}/{\bf Z})^{2g}$ of $Pic$.
Now observe that by a unitary transformation via
$$
U(P)\ =\ \exp\Bigl\{ \int_{P_0}^P A \Bigr\}
\eqno(37) $$
on a given spinor bundle we change the monodromies.
With a suitable $A\in J$ we can reach any point in $Pic$ and,
in particular, move to any other spin structure.
The NSR fermions couple to the abelian gauge field as in eq.~(37).
Hence, the sum over NSR spin structures is automatically contained in the
integration over the $U(1)$ moduli.
In fact, any change in NSR monodromies (jointly for all $\j^{i\m}$) can
be traded for a shift in $U(1)$ moduli space and, hence, cannot be physical.
This feature is not restricted to $g\ge 1$ but appears just as well
for the $n$-punctured sphere, i.e. in tree-level amplitudes.
Here, any (R,R) puncture can be turned into an (NS,NS) one,
since there are $n{-}1$ independent cycles and the sum of all twists has to
vanish. As a result, $\F$ and $\U$ states cannot be physically distinguished.
This is consistent with our observation that their correlators coincide.
However, at present we do not know an explicit spectral flow operator
implementing the $\F\leftrightarrow\U$ interchange on the level of the
vertex operators.

When the fermionic moduli are formally integrated out with a simple choice
of fermionic beltrami differentials,
the appropriate number of \pco s appear in the string measure.
Furthermore, we have to do the $U(1)$ moduli integration,
which adds to the measure $g$ insertions of a novel
`$U(1)$ projection operator' containing $\tilde b$ and $J_{\rm tot}$.
Its significance remains to be clarified.
Together with the usual $3g{-}3$ insertions of $b$ ghosts folded with
regular beltrami differentials and a single $\tilde c$ this completes the
measure for the remaining integration over the metric moduli of standard
Riemann surfaces.

\vglue.5in

\noindent{\bf Note added in 12/94:}

\noindent

N. Berkovits and C. Vafa pointed out to us the relevance of their recent
work~\cite{bv2,ber} where they imbed the $N{=}2$ string into a new $N{=}4$
topological string theory. This allows them to rewrite the critical
$N{=}2$ string $n$-point functions as correlators in the topological
theory, where their vanishing to all string loop orders can be proved,
with the known exceptions.
I would also like to mention a new preprint by H. L\"u and C. Pope~\cite{lp}
which overlaps with our results but takes a different viewpoint.

\vfill\eject


\begin{thebibliography}{99}
\bibitem{klp} S. V. Ketov, O. Lechtenfeld and A. J. Parkes,
{\it Twisting the $~N{=}2~$ string}, \\
Hannover preprint ITP-UH-24/93, hep-th/9312150 (corrected 8/94), \\
to appear in Phys. Rev. {\bf D}.
\bibitem{bkl} J. Bischoff, S. V. Ketov and O. Lechtenfeld, {\it The GSO
Projection, BRST \\
Cohomology and Picture-Changing in $N{=}2$ String Theory}, Hannover preprint \\
ITP-UH-05/94, hep-th/9406101 (revised 9/94), to appear in Nucl. Phys. {\bf B}.
\bibitem{aba} M. Ademollo, L. Brink, A. D'Adda, R. D'Auria, E. Napolitano,
S. Sciuto, E. Del Giudice, P. Di Vecchia, S. Ferrara, F. Gliozzi, R. Musto and
R. Pettorino, \pl{62}{76}{105}.
\bibitem{aba2} M. Ademollo, L. Brink, A. D'Adda, R. D'Auria, E. Napolitano,
S. Sciuto, E. Del Guidice, P. Di Vecchia, S. Ferrara, F. Gliozzi, R. Musto,
R. Pettorino and J. Schwarz, \np{111}{76}{77}.
\bibitem{ft} E. S. Fradkin and A. A. Tseytlin, \pl{106}{81}{63}.
\bibitem{ali} A. D'Adda and F. Lizzi, \pl{191}{87}{85}.
\bibitem{bien} J. Bienkowska, \pl{281}{92}{59}.
\bibitem{siegel} W. Siegel, \prl{69}{92}{1493}; \pr{46}{92}{3235},
{\it ibid.} {\bf D47} (1993) 2504 and 2512.
\bibitem{ov} H. Ooguri and C. Vafa, \mpl{5}{90}{1389}; \np{361}{91}{469}.
\bibitem{bv} N. Berkovits and C. Vafa, \mpl{9}{94}{653}.
\bibitem{ff} J. M. Figueroa-O'Farrill, \pl{321}{94}{344}.
\bibitem{op} N. Ohta and J. L. Petersen, \pl{325}{94}{67}.
\bibitem{bop} F. Bastianelli, N. Ohta and J. L. Petersen, \pl{327}{94}{35}.
\bibitem{ma} N. Markus, {\it A Tour through $N{=}2$ Strings}, Tel-Aviv preprint
TAUP-2002-92, November 1992, hep-th/9211059.
\bibitem{ket} S. V. Ketov, \cqg{10}{93}{1689}.
\bibitem{bsa} L. Brink and J. Schwarz, \np{121}{77}{285}.
\bibitem{mm} S. D. Mathur and S. Mukhi, \pr{36}{87}{465}; \np{302}{88}{130}.
\bibitem{fms} D. Friedan, E. Martinec and S. Shenker, \np{271}{86}{93}.
\bibitem{k-w} V. A. Kosteleck\'y, O. Lechtenfeld, W. Lerche, S. Samuel and
S. Watamura, \np{288}{87}{173}.
\bibitem{nst} A. Neveu, J. Schwarz and C. Thorn, \pl{35}{71}{529}.
\bibitem{gos} F. Gliozzi, D. Olive and J. Scherk, \np{122}{77}{253}.
\bibitem{wi} E. Witten, \np{276}{86}{291}.
\bibitem{we} C. Wendt, \np{314}{89}{209}.
\bibitem{ls1} O. Lechtenfeld and S. Samuel, \np{310}{88}{254}.
\bibitem{ls2} O. Lechtenfeld and S. Samuel, \pl{213}{88}{431}.
\bibitem{bgi} M. Bonini, E. Gava and R. Iengo, \mpl{6}{91}{795}.
\bibitem{bv2} N. Berkovits and C. Vafa, {\it $N{=}4$ Topological Strings},
Harvard and King's College preprint HUTP-94/A018 and KCL-TH-94-12,
July 1994, hep-th/9407190.
\bibitem{ber} N. Berkovits, {\it Vanishing Theorems for the Self-Dual
$N{=}2$ String}, S\~ao Paulo preprint IFUSP-P-1134, December 1994,
hep-th/9412179.
\bibitem{lp} H. L\"u  and C.N. Pope, {\it BRST Quantisation of the
$N{=}2$ String}, Texas and Trieste preprint CTP TAMU-62/94 and
SISSA-175/94/EP, Nov. 1994, hep-th/9411101.

\end{thebibliography}
\end{document}